\documentclass[aps,prl,twocolumn,showpacs]{revtex4-1}
\usepackage{natbib}
\usepackage{amsmath}
\usepackage{graphicx}
\usepackage{graphics}
\usepackage{subfigure}

\begin{document}

\title{Nonlocal Linear Compression of Two-Photon Time Interval Distribution}



\author{J.-S. Pan}
\author{X.-B. Zou}\email[]{xbz@ustc.edu.cn}
\author{Z.-Y. Zhou}
\author{D.-S. Ding}
\author{B.-S. Shi}\email[]{drshi@ustc.edu.cn}
\author{G.-C. Guo}
\affiliation{Key Laboratory of Quantum Information, University of Science and Technology of China, Hefei 230026, China}



\date{\today}
\begin{abstract}
We propose a linear compression technique for the time interval distribution of photon pairs. Using a partially frequency-entangled two-photon (TP) state with appropriate mean time width, the compressed TP time interval width can be kept in the minimum limit set by the phase modulation, and is independent of its initial width. As a result of this effect, ultra-narrow TP time interval distribution can be compressed with relatively slow phase modulators to decrease the damage of the phase-instability arising from the phase modulation process.
\end{abstract}


\maketitle


In some applications, including quantum positioning systems and quantum clock synchronization\cite{Bahder2004,Valencia2004,Giovannetti2001A}, the time intervals between two photons are employed as the precise timing information carriers. The experimental precision is directly linked to the width of the TP time interval distribution, i.e., the TP correlation time. However the temporal shape of photon pairs inherently has a certain width mainly owing to the bandwidth limit of the phase matching condition in the pair generation. Besides, the narrow temporal shape of photon pairs is very sensitive to the group velocity dispersion (GVD) in the propagation of the photons\cite{Valencia2002}. Therefore, in order to keep the TP correlation time in a narrow range, it is very important to develop a compression technology for TP time interval distribution.

The linear pulse compression technique with external phase modulation (PM) has been proposed for compressing optical pulses long ago\cite{Giodmaine1968}. In this technique, the frequency-broadening and the subsequent temporal compression effect of the pulses are obtain by the PM and dispersion compensation (DC) respectively. The phase modulator is adjusted to keep the truncation of the phase shift $\phi\left(t\right)$ imposing on the pulses to second order effective, which is the linear compression condition. This arrangement has avoided unnecessary complexity, since the frequency-broadening of the pulses mainly origins from the second-order term and the spectral phase arising from higher-order terms is hard to be compensated by GVD. The optimal compression ratio with respect to DC is given by $\sqrt{1+4\mu^{2}\tau^{4}}$, where $\tau$ is the pulse width and $\mu=\frac{d^{2}\phi\left(t\right)}{dt^{2}}$ is the so-called linear chirp rate. In order to obtain large $\mu$ in the effective range of the pulses, the pulses are usually arranged to centrally coincide with the peaks or valleys of $\phi\left(t\right)$. The effective varying speed of the phase modulators will sets a minimum limit for compressible pulse width\cite{Bartels2001}.

In principle, by independently compressing the signal and idler photons, one can access to the compression of the time interval distribution. However, on one hand, spontaneous parametric down-conversions (SPDC) in an aperiodically poled crystal has been demonstrated to generate chirped biphotons (i.e., strictly frequency-correlated photon pairs) with a large bandwidth\cite{Harris2007,Brida2009,Sensarn2010}. Bihotons with femtosecond (fs) correlation time are obtained by compensating the inhomogeneous spectral phase with GVD. On the other hand, the optimal achievable measuring precision of TP correlation is also in the fs range\cite{Donnell2009}. Therefore, significative compression technique for TP time interval distribution should able to compress fs pulses.
However, in the PM process of ultrashort pulses, the phase-instability between the probe radiation and the time-varying electric susceptibility is a ineluctable critical problem, especially in the current case that the TP radiation acts as precise timing information carriers, although it can be restrained with special arrangements in the single pulse case\cite{Bustard2010}.

In this letter, we propose a scheme to solve this problem and then access to the effective linear compression of TP time interval distribution.
We first identify the optimal compression of TP correlation time with a general chirp-dispersion interaction mechanism.
Under the optimal condition, we find the compression ratio takes a form similar to that in the case of single pulses. Merely the analogous role of the pulse width is replaced by the geometric mean of the TP correlation time and TP mean time width, which means that the compressed TP correlation time is independent of the initial one. As a result of this difference, ultrashort TP correlation time can be compressed with a slowly varying phase modulator to decrease the damage of the phase-instability in the PM process, by adjusting the trivial TP mean time width to match the PM to acquire enough modulation depth. An illustrative example is considered, in which the fs correlation time of the photon pairs generated by SPDC is compressed several times with with the phase modulators whose characteristic time is typically in the picosecond (ps) range. As a comparison, the single pulses with such narrow coincidence distribution width are almost immune to the PM.

Our theoretical analysis is divided into two steps: PM and DC (see Fig. 1(a)). We begin with a general TP state,
\begin{equation}\label{1}
|\psi\rangle\propto\int\int d\omega_{s}d\omega_{i}f\left(\omega_{s},\omega_{i}\right)|\Omega_{0}+\omega_{s},\Omega_{0}+\omega_{i}\rangle,
\end{equation}
where $\omega_{r}$ equals $\left(\Omega_{r}-\Omega_{0}\right)$ with the signal and idler frequency $\Omega_{r},r=s,i$, and the common central frequency $\Omega_{0}$. Since in general the photon pairs generated by SPDC really have common central frequency and their Sinc-function-type spectral shape can be approximatively replaced by a Gaussian one. We restrict our attentions to the TP states with symmetric TP frequency amplitude, i.e.,
\begin{equation}\label{2}
f\left(\omega_{s},\omega_{i}\right)=\exp\left(-\frac{1}{2}\tau_{1}^{2}\omega_{1}^{2}-\frac{1}{2}\tau_{2}^{2}\omega_{2}^{2}\right),
\end{equation}
where $\omega_{j}=\frac{\omega_{s}+\left(-1\right)^{j}\omega_{i}}{\sqrt{2}},j=1,2$ and the other parameters are constants. The frequency is
strictly correlated and anti-correlated at the zero and infinite limits of the ratio $R=\tau_{2}/\tau_{1}$ respectively.

As the time wave function of TP state, TP temporal amplitude is convenient to be employed to describe PM process. The TP temporal amplitude of the state $|\psi\rangle$ is defined as
\begin{equation}\label{3}
\mathcal{A}\left(t_{s},t_{i}\right)=\langle0|\hat{E}_{s}^{\left(+\right)}\left(t_{s}\right)\hat{E}_{i}^{\left(+\right)}\left(t_{i}\right)|\psi\rangle,
\end{equation}
where $\hat{E}_{r}^{\left(+\right)}\left(t_{r}\right)\propto \int d\Omega_{r}\hat{a}_{r}\left(\Omega_{r}\right)\exp\left(-i\Omega_{r}t_{r}\right),r=i,s$ is the electric field operator, and $\hat{a}_{r}\left(\Omega_{r}\right)$ is the annihilation operator. The TP correlation function of the state is given by $\left|\mathcal{A}\left(t_{s},t_{i}\right)\right|^{2}$. Under the linear compression condition, the phase shifts imposing on the signal and idler photons by phase modulators, $\phi_{r}\left(t_{r}\right)$, can be truncated to second-order terms. Then the PM process is equivalent to multiply the TP temporal amplitude by a factor, $\exp\left(i\frac{\mu_{s}t_{s}^{2}+\mu_{i}t_{i}^{2}}{2}\right)$, where $\mu_{r}=\frac{d^{2}\phi_{r}\left(t_{r}\right)}{dt_{r}^{2}}$ is the linear chirp rate. The lower-order terms of $\phi_{r}\left(t_{r}\right)$'s are ignored since they only contribute to constant phase and overall frequency shifts.

The chronocyclic Wigner function, whose definition for the state $|\psi\rangle$ is given by,
\begin{equation}\label{4}
\begin{split}
\mathcal{W}\left(t_{s},\omega_{s};t_{i},\omega_{i}\right)  =& \frac{1}{\left(2\pi\right)^{2}}\int\int d\tau_{s}d\tau_{i}\mathcal{A}^{*}\left(t_{s}-\frac{\tau_{s}}{2},t_{i}-\frac{\tau_{i}}{2}\right) \\
 & \times\mathcal{A}\left(t_{s}+\frac{\tau_{s}}{2},t_{i}+\frac{\tau_{i}}{2}\right)e^{i\omega_{s}\tau_{s}+i\omega_{i}\tau_{i}},
\end{split}
\end{equation}
is directly mapped into TP correlation function by its integration over the frequency plane. The dispersion process is described by multiplying the TP frequency amplitude with a factor $\exp\left(i\frac{\beta_{s}\omega_{s}^{2}+\beta_{i}\omega_{i}^{2}}{2}\right)$, where the dispersion parameter $\beta_{r}$ is the product of GVD and the propagation length of photons in the dispersive system, such as two dispersive media. Interestingly, the influences of the dispersion process can be manifested by the translations: $t_{r}\rightarrow t_{r}+\beta_{r}\omega_{r}, r=s,i$, on the time-frequency plane of the chronocyclic Wigner function\cite{Wasak2010}.

The TP correlation time $T_{c}$ can be defined as the root-mean square (r.m.s.) width of the relative time variable, $\left(t_{s}-t_{i}\right)$, under the TP correlation distribution\cite{Friberg1985}, and can be  measured by a coincidence measurement\cite{Friberg1985}, as shown in Fig. 1(a). Unless otherwise specified, all time widthes in this letter refer to the r.m.s. type. For the state $|\psi\rangle$ in Eq. (1)-(2), the initial TP correlation time $T_{c,i}$ is $2\tau_{1}$. Then the TP correlation time after the PM and dispersion process, $T_{c,f}$, is given by the width of the cross variable, $\left(t_{s}-\beta_{s}\omega_{s}-t_{i}+\beta_{i}\omega_{i}\right)$, calculated before the dispersion processes. After some calculations, we derive the form of $T_{c,f}$,
\begin{widetext}
\begin{equation}\label{5}
T_{c,f}=2\sqrt{\tau_{1}^{2}+\frac{1}{2}\sum_{j=1}^{2}\beta_{j}^{2}\left[\frac{1}{\tau_{3-j}^{2}}+\frac{1}{2}\left(\mu_{1}^{2}\tau_{j}^{2}+\mu_{2}^{2}\tau_{3-j}^{2}\right)\right]-\beta_{1}\mu_{1}\tau_{1}^{2}-\beta_{2}\mu_{2}\tau_{1}^{2}+\frac{1}{2}\beta_{1}\beta_{2}\mu_{1}\mu_{2}\left(\tau_{1}^{2}+\tau_{2}^{2}\right)},
\end{equation}
\end{widetext}
where $s_{j}=\frac{s_{s}-\left(-1\right)^{j}s_{i}}{\sqrt{2}},s=\mu,\beta$.

In principle, the dispersion parameters have on restrictive conditions. We first consider the maximal value of $T_{c,f}$, $T_{c,f,d}$, with respect to the optimal CD for arbitrary PM.
According to Eq. (4), $T_{c,f}$ relies on a binary quadratic equation of $\beta_{j}$'s. The first derivative of the function is set to zero to find the maximal value of $T_{c,f}$. Here the compression ratio, $T_{c,i}/T_{c,f,d}$, takes the form,
\begin{equation}\label{6}
 Rc_{d}=\sqrt{\frac{\left(1+R^{2}\right)^{2}\left(x+y\right)^{2}+4\left(1-R^{2}xy\right)^{2}}{R^{2}\left(1+R^{2}\right)\left(x+y\right)^{2}+4\left(1-R^{2}xy\right)}},
\end{equation}
where $x=\mu_{s}\tau_{1}^{2}$ and $y=\mu_{i}\tau_{1}^{2}$ are the dimensionless linear chirp rates.

The inverse of $Rc_{d}$, as a function of $x$ and $y$, is depicted in Fig. 1(b)-(d) with respect to $R=1$ (b), $R=4$ (c), $R=16$ (d). From the plots, we find the values of $1/Rc_{d}$ always drop fastest along the line $x+y=0$. Besides, the relative falling speed along this line increases with $R$ increasing. In fact, as $R\gg1$, $Rc_{d}$ approximates to $\sqrt{1+4x^{2}y^{2}/\left(x+y\right)^{2}}$, which amounts to the compression ratio even plunges near this line. It means that the optimal PM process is to make the linear chirp rates of the photon pairs have opposing values. With this arrangement, i.e., $\mu_{s}=-\mu_{i}$,
the optimal compression ratio $Rc_{opt}$ is given by,
\begin{equation}\label{7}
Rc_{opt}=\sqrt{1+\mu^{2}\tau_{1}^{2}\tau_{2}^{2}}.
\end{equation}
The dispersion parameters need be set as, $\beta_{s}=-\beta_{i}=\beta=\frac{\mu\tau_{1}^{2}\tau_{2}^{2}}{\mu^{2}\tau_{1}^{2}\tau_{2}^{2}+1}$. The dispersion completely compensates the spectral phase introduced in the PM process.

The interference dip width of HOM interference is also usually used to characterize the width of TP time interval distribution\cite{Hong1987}. The normalized coincidence counting rate $R_{n}(\tau)$ can be written as
\begin{equation}\label{8}
R_{n}\left(\tau\right)=1-\xi\int dt_{1}dt_{2}\mathcal{A}\left(t_{1},t_{2}\right)\mathcal{A}^{\ast}\left(\sqrt{2}\tau-t_{1},t_{2}\right),
\end{equation}
where $\mathcal{A}\left(t_{1},t_{2}\right)$ is an alternative form of $\mathcal{A}\left(t_{s},t_{i}\right)$ with the cross variables $t_{j}=\frac{t_{s}+\left(-1\right)^{j}t_{i}}{\sqrt{2}},j=1,2$, $\tau$ is the relative delay between signal and idler paths. The dip width of $R_{n}(\tau)$, which is also referred as to TP coherence time, is related to the overlap of the TP temporal amplitude and the time reversal of its conjugate. The general analysis of the linear compression of TP coherence time is complicated, but its compression ratio under the optimal compression condition referred above is also given by Eq. (6). Then the subsequent discussion for TP correlation are also suitable for it.

\begin{figure}
\centering
\includegraphics[width=3.3in]{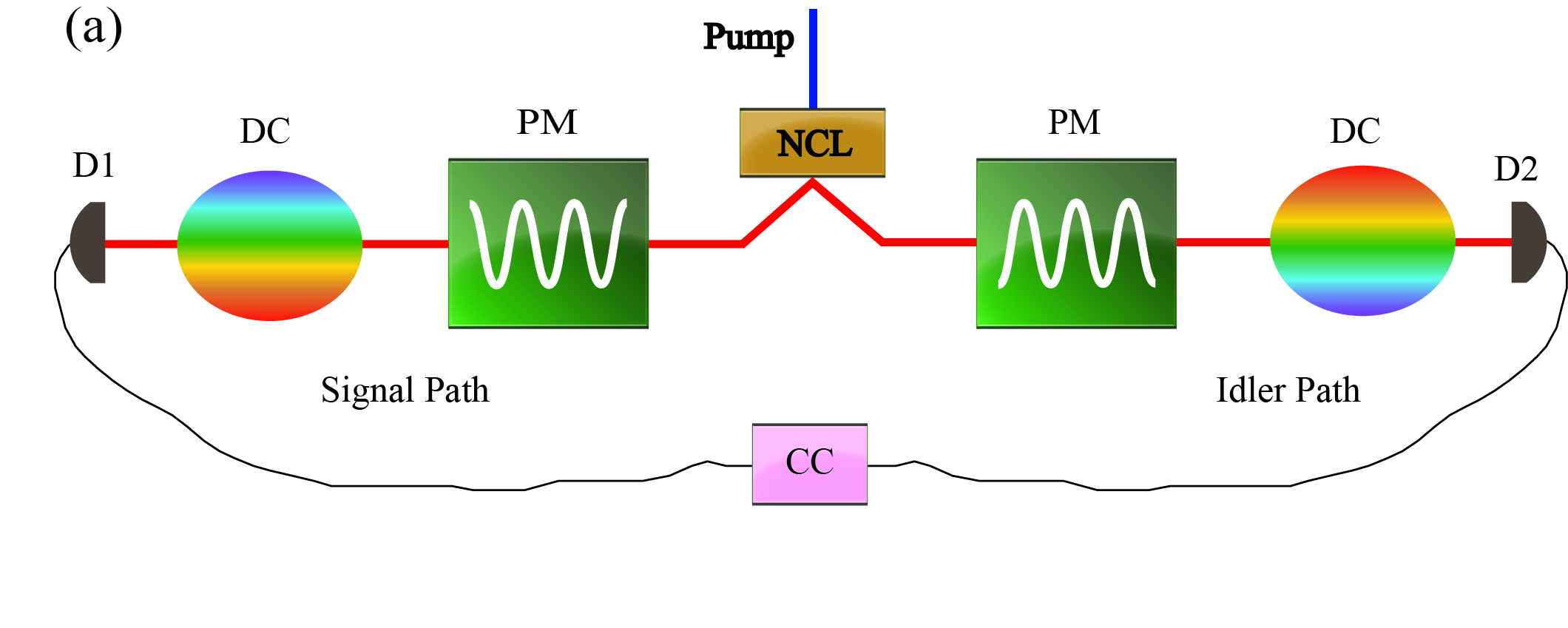}
\includegraphics[width=1in]{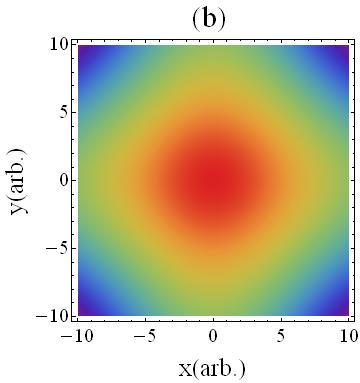}
\includegraphics[width=1in]{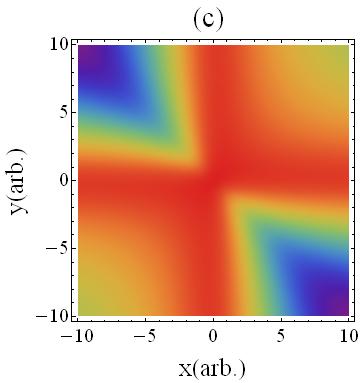}
\includegraphics[width=1.18in]{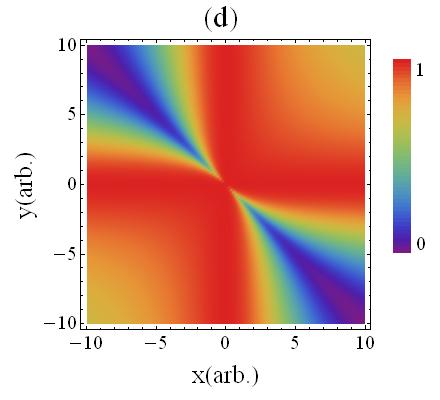}

\caption{\label{9} (color online). (a) A schematic diagram of the linear compression of photon pairs. PM (phase modulation) and DC (dispersion compensation) are two main steps in our arrangement. For intuitiveness, we plot the source as SPDC in a nonlinear crystal (NLC) and measurement system as a standard TP coincidence measurement, where CC denotes a coincidence counter, D1 and D2 denotes detectors. As shown in the plot, the chirp and dispersion parameters are opposing in the two paths. (b)-(d): The inverse of the compression ratio, $1/Rc_{d}$, as a function of $x$ and $y$, with $R=1$ (b), $R=4$ (c) and $R=16$ (d). It's shown that the values of $1/Rc_{d}$ always drops fastest along the line $x+y=0$. Besides, the relative falling speed along this line is increase with $R$.}
\end{figure}

In the separable case that $\tau_{1}=\tau_{2}=\sqrt{2}\tau$, the optimal compression ratio $Rc_{opt}$ transfers to that of the coincidence distribution width of two independent pulses with a width $\tau$, which is the classical analogs of TP correlation time\cite{Franson1992}. Therefore, the classical linear compression of two independent pulses is a special case of the linear compression of entangled photon pairs. In the single pulse case, the compressed pulse width approximates to $1/2\mu\tau$, which is anti-proportional to the initial pulse width $\tau$. It means that the attainable compressed shape of a long pulse is narrower than that of a short pulse\cite{Bartels2001}. Besides, when the phase modulators vary so slow that the chirp rate $\mu$ is relatively small compared with $1/2\tau^2$, the pulses can't be effectively compressed.

In contrast, the optimal compressed TP correlation time approximates to $1/\mu\tau_{2}$, which is independent of the initial width $2\tau_{1}$. This makes our scheme very suitable for keeping the TP time interval width in a narrow range. Besides, even if when we employ slowly varying phase modulators so that the chirp rate $\mu$ is relatively small compared with $1/\tau_{1}^{2}$, the compression ratio $Rc_{opt}$ may still has a considerable value if $\tau_{2}$ is selected to be far larger then $\tau_{1}$. Since if the interval between the time-origin of the probe TP radiation and the PM process randomly shifts $\delta$, the phase and frequency of the electric field operator $\hat{E}_{r}^{\left(+\right)}\left(t_{r}\right)$ shift $\mu\delta
 $ and $\mu\delta^2/2$ respectively. The use of slowly varying phase modulators will decrease the damage of the phase-instability. Certainly, since when $\tau_{2}$ far larger than $\tau_{1}$, the TP temporalamplitude has a long narrow shape along the line $t_{s}-t_{i}=0$, and the amplitude of the signal and idler radiation has a width approximating $\tau_{2}$.
The phase modulators should have a long characteristic time to ensure the effective modulation depth, $\mu \tau_{2}^2/2$, has a value larger compared with $\mu/\tau_{1}^2$, when $\tau_{2}$ accesses to the limit set by the linear compression condition.
Tt means that the compressible TP correlation for a phase modulator has a certain limit.

The phase variation generated by the phase modulators meeting the condition $\mu_{s}=-\mu_{i}$, can be shown by the function,
$\cos\left(\mu t_{1}t_{2}\right)$, which is potted as the background of Fig. 2(a). The phase shift of an entangled TP temporal amplitude, $\exp\left(-\frac{t_{1}^{2}}{2\tau_{1}^{2}}-\frac{t_{2}^{2}}{2\tau_{2}^{2}}\right)$, with $\tau_{2}>\tau_{1}$,
is shown by the fluctuation density in the solid elliptical ring.
That of a separable TP temporal amplitude with similar shape is shown by the dashed ring.
For the entangled one, no matter how small $\mu$ is or how narrow the width with respect to the cross variable $t_{1}$ is,
we always can obtain considerable modulation in the relative time direction by increasing $\tau_{2}$.
But for the separable one, the phase shifts along the two main axes are independent.
This suggests the effect referred in the above paragraph relies on the non-locality of the TP state.
Comparing with nonlocal dispersion cancellation and nonlocal PM effects, which are based on strict frequency-correlation\cite{Franson1992,Harris2008}, the phenomenon here idicates a applicable nonlocal effect for the chirp-dispersion response of photon pairs based on partially frequency-correlation.

\begin{figure}
\centering
\includegraphics[width=1.2in]{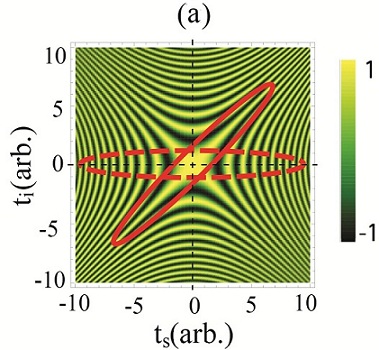}
\includegraphics[width=2.1in]{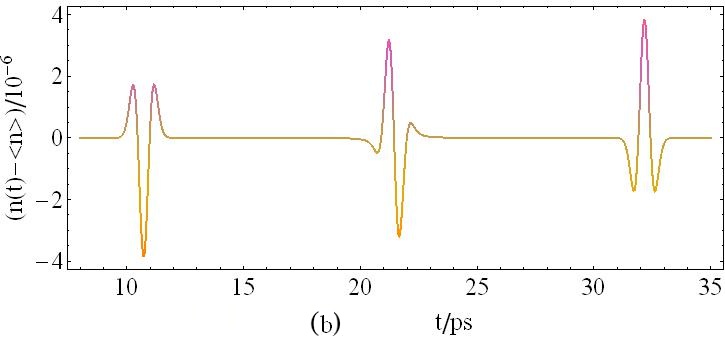}
\includegraphics[width=1.2in]{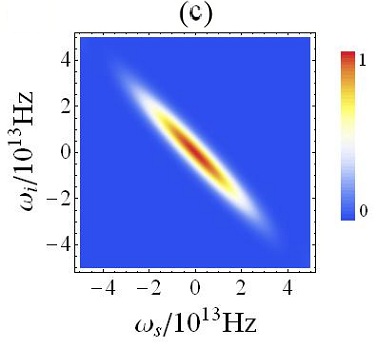}
\includegraphics[width=2.1in]{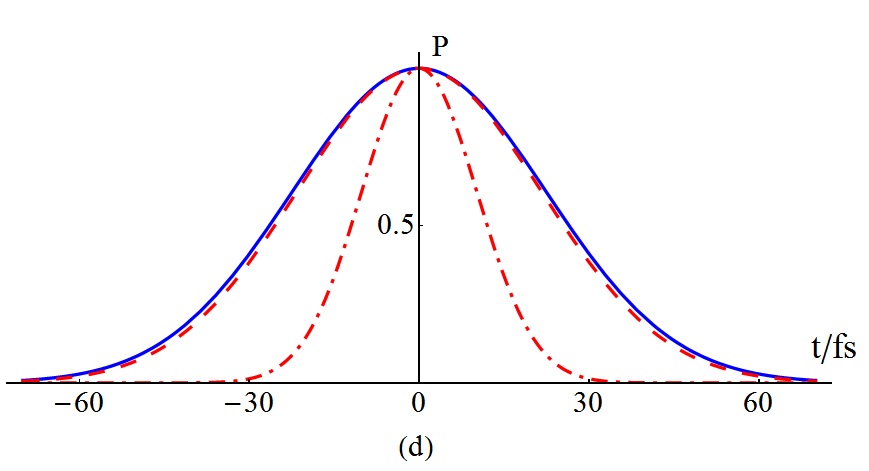}
\caption{
\label{9} (color online).(a) The background shows the function, $\cos\left(\mu t_{1}t_{2}\right)$. The red solid and dashed elliptical rings indicate the effective ranges of an entangled TP temporal amplitude and a separable one respectively. (b) The time-varying refractive index $n\left(t\right)$ of the $\text{CO}_{2}$ gas. The time-origin is decided by the pump pulses. (c) The joint spectral intensity of the photon pairs. (d) The blue solid line indicates the common initial shape of the conditional distribution of $t_{1}$ for the photon pairs and the relative intensity distribution of the single pulses. The red doted-dashed line and red dashed line respectively indicate their compressed shapes.
}
\end{figure}

We analyse a practical example with the optimal compression condition.
The type-II SPDC in a $1\text{mm}$ long KDP crystal that pumped by $0.22\text{ps}$ pulses is used to generate entangled photon pairs\cite{Giuseppe1997}.
The joint spectral shape, which is shown in Fig. 2(c), is close to that characterized by Eq. (2).
The analogous ratio of $R$ is about $7$. The TP correlation time $T_{c}$ is about $63.3\text{fs}$.
The rotationally exited $\text{CO}_{2}$ molecules are employed as phase modulators\cite{Bartels2001}.
The probe-pump geometric arrangements and the parametric settings are identical with those in the Ref.\cite{Bartels2001}.
The calculation of the time-varying refractive index is based on the calculation of that of the linear molecular gases in the Ref. \cite{Lin1971}.
The characteristic time of the modulators is typically in ps range, as shown in Fig. 2(b).
The signal and idler photons are adjusted to centrally coincide with the first- and third-partial rotational revivals to simulate the optimal PM process.
In order to generate tunable dispersion parameters, one can use the setup proposed in the Ref. \cite{Donnell2011}. In our calculation, the dispersion parameters are set to $\pm2400 \text{fs}^2$.
The distribution of $t_{1}$ with $t_{2}=0$, whose width is about $T_{c}/\sqrt{2}$, is depicted in Fig. 2(d) with the blue solid line.
The compressed distribution corresponds to the red doted-dashed line, which means that that $T_{c}$ is compressed about $2.3$ times.
As a comparison, we consider the optimal compression of the pulses whose coincidence distribution width is identical with $T_{c}$.
The compressed intensity distribution of the pulses are shown by the red dashed line.
From the plot, we find the pulses have nearly no change.

In conclusion, we develop a technique for linearly compressing TP time interval distribution.
It's shown that, under certain conditions, the modulation of TP relative time variable via the PM can be non-locally enhanced by stretching TP mean time distribution.
Therefore, using entangled TP state with appropriate mean time width, the compressed TP time interval width can be kept in the minimum limit set by the phase modulation.
On one hand, this effect enables us to use relatively slow phase modulators in the compression of ultra-narrow TP time interval distribution to decrease the damage of the phase-instability arising from the PM process.
On the other hand, it may provide new insights into the non-locality of TP state.
Our work may open up a new road to the application in quantum physics for PM techniques.

\begin{acknowledgments}
This work was supported by the National Natural Science Foundation of China (Grants No. 11174271, 61275115, 10874171 and 11274295), the National Fundamental Research Program of China (Grant No. 2011CB00200), and the Innovation fund from CAS, Program for NCET. Correspondence and requests for materials should be addressed to BSS.
\end{acknowledgments}

\bibliography{main}
\end{document}